\documentclass[12pt]{article}
\usepackage{hyperref}
\usepackage{cite}
\usepackage[dvips]{color}
\usepackage{epsfig}
\usepackage{amsmath}
\usepackage{graphicx}
\usepackage{hyperref}
\usepackage{footnote}
\makesavenoteenv{minipage}
\renewcommand{\title}[1]{%
    \bigskip%
    \begin{center}%
    \Large\bf #1%
    \end{center}%
    \vskip .2in}

\renewcommand{\author}[1]{%
    {\begin{center}
    #1
    \end{center}}}

\begin{document}
\title{\bf  Thermal Fluctuations in a Hyperscaling Violation Background}
\author{{Behnam Pourhassan$^{a}$\footnote{Email: b.pourhassan@du.ac.ir},
Mir Faizal$^{b,c}$\footnote{Email: f2mir@uwaterloo.ca}, Sudhaker Upadhyay,$^{d}$\footnote{E-mail: sudhakerupadhyay@gmail.com} and Lina Al Asfar$^{e}$\footnote{Email: linana343@outlook.com}}\\
$^{a}${\small {\em School of Physics, Damghan University, Damghan, 3671641167, Iran}}\\
$^{b}${\small {\em Irving K. Barber School of Arts and Sciences,  University of British Columbia-Okanagan}}\\
{\small {\em Kelowna, British Columbia V1V 1V7, Canada}}\\
$^{c}${\small {\em Department of Physics and Astronomy, University of Lethbridge, Lethbridge, Alberta, T1K 3M4, Canada}}\\
$^{d}${\small {\em Centre for Theoretical Studies, Indian Institute of Technology Kharagpur,  Kharagpur-721302,  India}}\\
$^{e}${\small {\em Laboratoire de Physique Corpusculaire de Clermont-Ferrand,}}\\
{\small {\em Universit\'{e} Blaise Pascal, 24 Avenue des Landais F-63177 Aubi\`{e}re Cedex, France}}}

\begin{abstract}
\noindent In this paper, we  study the effect of thermal fluctuations on the
thermodynamics of
a black geometry with hyperscaling violation.
These thermal fluctuations in the thermodynamics of this system  are produced
from quantum corrections of geometry  describing this system.
We   discuss  the stability of
this system using    specific heat   and the entire Hessian
matrix of the free energy. We will  analyze the effects of thermal fluctuations on
the stability of this system. We also analyze the effects of thermal fluctuations on the
criticality of the Hyperscaling violation background.

\end{abstract}
\section{Introduction}
It is important to associate an entropy with black holes to prevent the violation
of the second law of thermodynamics. This is because
if  black holes were not maximum entropy objects, then the entropy of the universe
would spontaneously reduce, whenever an object with a finite entropy crossed the horizon.
So, black holes are maximum entropy objects, and they have more entropy
than any other object with   the same volume \cite{1, 1a, 2, 4, 4a}.
The scaling of this  maximum entropy  with the area  of the  horizon has led to the
development of the holographic principle \cite{5, 5a}.
The holographic principle equates the degrees of freedom in any region
of space with the degrees of
freedom on the boundary of that region.

The holographic principle is expected to be corrected near Planck scale, as quantum gravity
corrections modify the manifold structure of space-time at Planck scale \cite{6, 6a}.
As the holographic principle was motivated from the entropy-area relation, it can be argued
that the quantum gravity corrections would also modify this entropy-area relation.
Now for a black hole with area $A$ and entropy $S_0$,
original entropy-area relation in natural unites is given by $S_0= A/4$.
However, the corrected entropy-area relation can be written as
$S=  S_0 + \alpha \log A+  \gamma_1 A^{-1} + \gamma_2 A^{-2} \cdots$, where $\alpha,
\gamma_1, \gamma_2 \cdots$,
are coefficients which depend on the details of the model. The general structure
of the corrections and their dependence on the area is a universal feature, and it occurs in
  almost all approaches to quantum gravity.
The  corrections to the thermodynamics
  of black holes have been studied using the    non-perturbative quantum  general
relativity  \cite{1z}. In this formalism, the conformal blocks of a well
defined  conformal field theory were used to study the behavior of the
density of states of a black hole. The quantum corrections to the thermodynamics of a black hole
has also been studied using the Cardy formula  \cite{card}.  The corrected  thermodynamics of
a black hole has also been studied by analyzing the effect of
  matter fields surrounding a black hole
\cite{other, other0, other1}. Such corrections have the general feature that they are
represented by a logarithmic function of area.

As string theory is one of the most important approaches to
quantum gravity, it is very important to understand the effects of quantum corrections
produced by string theoretical effects. In fact, the  corrections  produced by string theoretical
effects on  the thermodynamics of a black hole have been studied, and it has been observed
that they produce the same general form of the corrections as are produced from
other approaches to quantum gravity    \cite{solo1, solo2, solo4, solo5}.
The  corrections to the thermodynamics of a  dilatonic black hole have also been studied,
and they have again been observed to have the same universal form \cite{jy}. The partition
function of a black hole  has  also been used to analyze the corrections to the thermodynamics
of a black hole \cite{bss}. Another universal feature of almost all theories of quantum gravity
is the existence of a minimum measurable length scale, and this has motivated the generalized
uncertainty principle, which in turn has been used to study corrections to the
thermodynamics of the black holes  \cite{mi, r1}.
It has been demonstrated that the black hole thermodynamics corrected by the generalized
uncertainty principle again has the same universal form as produced by other approaches to quantum
gravity.

In fact, the universality of this correction can be argued from purely thermodynamics arguments.
This is because in the  Jacobson formalism, the  Einstein
equations  are thermodynamics identities  \cite{z12j, jz12}.
So, space-time geometry is am emergent property from the thermodynamics.
Thus, a thermal fluctuation in thermodynamics would produce a quantum fluctuation
in the geometry of space-time.
In fact, it has been demonstrated that
the thermal fluctuations correct the thermodynamics of black holes, and this corrected
thermodynamics
has the same universal form as expected from the quantum gravitational effects
\cite{l1, SPR, more}.
As the coefficient of   quantum gravity corrections depend on the details of the model,
we will keep such a coefficient as a constant. Thus, we will analyze the corrections produced
by thermal fluctuations on the near horizon geometry of a hyperscaling violating background,
with variable coefficients.

It may be noted that such thermal corrections have been studied for various
different black geometries.
Such   corrections   have    been studied for  G\"{o}del black hole \cite{godel}.
The such corrections for an AdS charged black hole has been  studied,
and it has been observed that the thermodynamics of this AdS black hole is modified
by thermal fluctuations \cite{1503.07418}. The effect of thermal fluctuations on the
thermodynamics  for a black Saturn have also been studied   \cite{1505.02373}. It has
been demonstrated that the thermal fluctuations do not have any major effect on the
 stability of the black Saturn. The  thermal fluctuations for a
 modified Hayward black hole have
been studied, and it has been demonstrated that
such thermal fluctuations reduce the pressure
and internal energy of such a black hole \cite{1603.01457}.
 The effect of thermal fluctuations on the thermodynamics of a
 charged dilatonic black Saturn has been also studied
 \cite{1605.00924}. It was observed that  the thermal fluctuations can be studied
 either using a conformal field theory or by analyzing the fluctuations in the energy
 of this system. However, it has been demonstrated that the
 fluctuations in the energy and the conformal field theory produce the same results
 for a charged dilatonic black Saturn. The effects of thermal fluctuations on the
 thermodynamics of a small singly
 spinning Kerr-AdS black hole have also been studied
  \cite{NPB}.  As dumb holes are analogs black hole like solutions, it is possible
  to study such effects for dumb holes. In fact, the effects of thermal fluctuations
  on the thermodynamics of dumb holes has been studied   \cite{Annals}.
The thermal fluctuations can affect the critical behaviors of black holes.
Such corrections to a
 dyonic charged anti-de
Sitter black hole have been studied, and it has been observed that such a corrected
solution also describes a
van der Waals fluid \cite{PRD}.
It is also expected that such corrections to the solutions in AdS
can be used to study the modifications to the
  quark-gluon
plasma, and this can be done  using the AdS/CFT correspondence \cite{JHEP,CAN,EPJC2,G}.
In fact, it has been demonstrated that such corrections can produce interesting
modifications to the ratio between the  viscosity to entropy of such a system
\cite{EPJC}.

Now, as this form of the corrections to the thermodynamics is universal,
we will analyze the effects
of such corrections on the thermodynamics of a near horizon geometry with hyperscaling
violation. Hyperscaling violating backgrounds are interesting geometries,
and have been used to study interesting physical systems
\cite{hyper001,hyper002,hyper003}. The
possible boundary conditions of scalar fields in a hyperscaling violation
 geometry has been studied  \cite{hyper004}.
Such backgrounds are also interesting for holographic models \cite{hyper005},
such as  holographic superconductor \cite{hyper006}. Singularities in Hyperscaling
violating space-times have also been investigated \cite{hyper007}.
The analytic solution of a Vaidya-charged black
hole with a hyperscaling violating factor (in an Einstein-Maxwell-dilaton model \cite{EMd})
has been obtained   \cite{hyper008}. The
 hyperscaling violating solutions in
  Einstein-Maxwell-scalar theory have also been studied \cite{001,002,003}.
Thermalization of mutual information in hyperscaling violating backgrounds has been
analyzed, and it has been observed  that the dynamical
exponent is  important for understand  mutual information in such a system  \cite{hyper009}.
Entanglement temperature \cite{hyperT} and entanglement entropy \cite{hyperS}
with hyperscaling violating backgrounds are also been investigated.

The a black brane geometry with
hyperscaling violating backgrounds can also be constructed, and it is possible to
study  the thermodynamics of such a system.
In fact, the thermodynamics of
nonlinear charged Lifshitz black branes with hyperscaling violation has been discussed
\cite{hyperTerm}. In this system, the effect of a nonlinear electromagnetic field
on the hyperscaling violating Lifshitz
black branes   has been studied. It may be noted that as the
hyperscaling violating geometries
have been used for analyzing various different physical systems
\cite{hyper3,hyper4,hyper5, hyper6},
 it would be interesting to analyze the effects of quantum corrections
on such a hyperscaling violating geometry.
In this paper, we will analyze the corrections produced by the thermal fluctuations on the
near horizon geometry of a black brane \cite{hyper0000}.
  The Null-Melvin-Twist
and KK reduction have been used to analyze a hyperscaling violating geometry constructed
from a black brane \cite{hyper1, hyper2}. As this is an
  interesting geometry, and it
is both important and interesting to analyze the effects of thermal fluctuations
on such a geometry.
Thus, we will analyze the effects of thermal fluctuations on the
thermodynamics of this geometry. It will be observed that such correction terms  can have
very interesting effects on this system.

\section{Hyperscaling Violating Background}
In this paper, we will analyze the corrections to the thermodynamics of a near horizon geometry
of a black brane with hyperscaling violating.
Such a geometry is described by the following metric  \cite{hyper2},
\begin{eqnarray}\label{metric04}
ds^2_{d+2}&=&\left(\frac{r}{R}\right)^2\left(\frac{r_F}{r}\right)^{2\theta/d}\left(-\left(\frac{R^2}{r}\right)^{-2(z-1)}fdt^2+dy^2+dx^2_i\right)\nonumber\\
&+&\left(\frac{R}{r}\right)^2\left(\frac{r_F}{r}\right)^{2\theta/d}\left(\frac{dr^2}{f}+R^2d\Omega^2_{d+2}\right),
\end{eqnarray}
where $R$ is the AdS scale, $x_i=(x_1,x_2)$,   $r_F$
is a scale  describing this system \cite{hyper2}, and
\begin{equation}\label{metric05}
f=1-\left(\frac{r_h}{r}\right)^{d+z-\theta},
\end{equation}
with  $r=r_h$ as the radius of the horizon.
Now we can study the   finite temperature effects of hyperscaling
violation using the  condition, $r_F<r_h$. It may be noted that this analysis
can be simplified by   setting $z=1$, and this corresponds to having
an asymptotically AdS space-time \cite{hyper1}.  The   Null-Melvin-Twist and
KK reduction can now be used to obtain the following metric,
\begin{eqnarray}\label{metric06}
ds^2_{d+2} &=& K^{-2/3}\left(\frac{r}{R}\right)^2M\left[-(1+b^2r^2M^2)fdt^2-2b^2r^2fM^2dtdy+(1-b^2r^2fM^2)dy^2+Kdx^2_i\right]\nonumber\\
&+& K^{1/3}M\left(\frac{R}{r}\right)^2f^{-1}dr^2,\nonumber\\
\phi &=&-\frac{1}{2}\ln K,\nonumber\\
A&=&\frac{M^2}{K}\left(\frac{r}{R}\right)^2b(fdt+dy),
\end{eqnarray}
where $A$ is one-form field, $\phi$ is dilaton,  $
K=1-(f-1)br^2M^2 $,    $b$ is a
  free parameter (inverse of length), and
\begin{equation}
M=\left(\frac{r_F}{r}\right)^{(2\theta/d)}.
\end{equation}
This can be simplified  using the following light-cone coordinates,
\begin{eqnarray}\label{light-cone}
x^{+}&=&bR(t+y),\nonumber\\
x^{-}&=&\frac{1}{2bR}(t-y).
\end{eqnarray}
Thus, we can obtain the following  scaled extremal metric,
\begin{eqnarray}\label{metric10}
ds^2_{d+2} &=&
\left(\frac{r}{R}\right)^2\left(\frac{r_F}{r}\right)^{(2\theta/d)}
\left[\left(\frac{r}{R}\right)^2\left(\frac{r_F}{r}\right)^{(4\theta/d)}H^2_B{dx^{+}}^2-2iH_Bdx^{+}dx^{-}
+G^2_B{dx_{i}}^2\right]\nonumber\\
&+& \left(\frac{R}{r}\right)^2\left(\frac{r_F}{r}\right)^{(2\theta/d)}dr^2,
\end{eqnarray}
where $G_B = K\left(r_B\right)^{1/6}$, and
\begin{eqnarray}\label{func}
H_B &=& \left[K(r_B)^{-2/3}\left(\frac{f(r_B)-1}{(2bR)^2}+
{\left(\frac{r_B}{R}\right)}^2\left(\frac{r_F}{r_B}\right)^{(4\theta/d)}f(r_B)\right)\right]^{1/2}
\left(\frac{r_B}{R}\right)^{-1}\left(\frac{r_F}{r_B}\right)^{(-2\theta/d)}.\nonumber\\
\end{eqnarray}
In order to write metric (\ref{metric10}),
we analytically continue $x^{+}$ to $ix^{+}$, and assume the system
is inside a box by using the cut off $r=r_{B}$.
Here, the  finite cutoff $r_B$ is larger than the scale $R$.
The metric (\ref{metric10}),  can be obtained from the equations of motion that
follow from the action \cite{action},
\begin{eqnarray}
S=\frac{1}{16\pi G_{d+2}}\int dx^{d+2}\sqrt{-g}\left[{\cal R}-\frac{4}{3}\partial_\mu \phi\partial^\mu \phi -\frac{1}{4}R^2e^{-8\pi/3}F_{\mu\nu}F^{\mu\nu}-4A_\mu A^\mu -4e^{2\pi /3}\frac{(e^{2\pi}-4)}{R^2}  \right],
\end{eqnarray}
where $G_{d+2}$,  ${cal R}$ and $F_{\mu\nu}$ are the $(d+2)$ dimensional Newton constant,
the scalar curvature and field-strength tensor, respectively.
\section{ Corrected Thermodynamics}
In this section, we will analyze the effects of thermal fluctuations on the thermodynamics
of such a system. So, first we will obtain the thermodynamic quantities like
 the  entropy, the
free energy and the  temperature  for such a geometry \cite{hyper1}.
The Hawking temperature of this geometry can be expressed as
\begin{equation}\label{temp}
T_H=\frac{d+1-\theta}{4}\frac{r_{h}}{\pi b R^{3}},
\end{equation}
and the  local temperature for this geometry is given by
\begin{equation}\label{ltemp}
T=\frac{T_{H}}{\sqrt{f}},
\end{equation}
where $f$ given by the equation (\ref{metric05}).
The entropy of this geometry is given by
\begin{equation}\label{entropy}
S_{0}=\frac{4\pi b(d-\theta)(4d-3\theta)}{d^{2}R^{2}(d+1-\theta)}r_{h}^{\frac{3(d-\theta)}{d}}.
\end{equation}
Here the original entropy is denoted by  $S_0$, and the
 entropy corrected by thermal fluctuations will be denoted by
 $S$.
Since the above metric
  is stationary but not static and  describes the black brane rotating in the compactified
$x^-$ direction, therefore it is natural to interpret the angular momentum and
velocity as the charge
and the conjugate chemical potential of the system, respectively.
So, the  constant chemical potential for this system can be written as  \cite{hyper1}
\begin{equation}\label{chempot}
\mu=\frac{1}{2b^{2}R^{2}}.
\end{equation}
The  Helmholtz free energy of this system can be written as
\begin{equation}\label{Helmholtz}
F=E-TS,
\end{equation}
where $E$ is the internal energy of this system.

These thermodynamic quantities are calculated by neglecting the  effects of thermal
fluctuations on the thermodynamics of this system. This is valid as long as the
temperature is sufficiently small. However, as the temperature of the black geometry increases,
due to the decrease in the radius of the black geometry, we have to also consider the effects
of thermal fluctuations. If the thermal fluctuations are still small enough, then they
can be analyzed as a perturbation around the equilibrium state. These
thermal fluctuations for a black geometry can be expressed as
\cite{l1, SPR, more}
\begin{equation}\label{9}
S = S_{0} -\alpha \ln (S_{0} T^{2})+\frac{\gamma_{1}}{S_{0}}+\frac{\gamma_{2}}{S_{0}^{2}}+\cdots,
\end{equation}
where $\alpha$, $\gamma_{1}, \gamma_{2} \cdots $,
are correction parameters which depend on
details of the model being studied.
Using  (\ref{entropy}) and (\ref{ltemp}), the explicit form of $S$ can be written as
\begin{eqnarray}
S &=&\frac{4\pi b(d-\theta)(4d-3\theta)}{d^{2}R^{2}(d+1-\theta)}r_{h}^{\frac{3(d-\theta)}{d}}
-\alpha \log \left[\frac{4\pi b(d-\theta)(4d-3\theta)}{d^{2}R^{2}(d+1-\theta)}r_{h}^{\frac{3(d-\theta)}{d}}\right] \nonumber\\
&-&2\alpha \log \left[ \frac{(d+1-\theta)r_{h}}{4\pi b R^{3}} \frac{1}{\sqrt{1-\left(\frac{r_h}{r}\right)^{d+z-\theta}}}\right]+\gamma_1
\frac{d^{2}R^{2}(d+1-\theta)}{4\pi b(d-\theta)(4d-3\theta)}r_{h}^{-\frac{3(d-\theta)}{d}}\nonumber\\
&+&\gamma_2\frac{d^{4}R^{4}(d+1-\theta)^2}{16\pi^2 b^2(d-\theta)^2(4d-3\theta)^2}
r_{h}^{- \frac{6(d-\theta)}{d}}.
\end{eqnarray}
As their value of these coefficients depends  on the specifics of the model,
we will keep these values as a variable, in this paper.
We will now neglect  higher order corrections ($\gamma_{3}=0$) to this system.
So, the corrected  internal energy, which can be calculated from the  definition
$E=\int{TdS}$ is given by
\begin{eqnarray}\label{E}
E &=&\frac{(d+1-\theta) }{4\pi b}\frac{r_h}{\sqrt{1-\left(\frac{r_h}{r}\right)^{d+z-\theta}}}
\left[\frac{8\pi b(d-\theta)(4d-3\theta)}{5d^3 R^5 (d+1-\theta)}r_h^{\frac{3(d-\theta)}{d}}
-\frac{8\alpha}{3R^3} \right.\nonumber\\
&-&\left. \gamma_1\frac{d^2(d+1-\theta)}{2\pi bR(d-\theta)(4d-3\theta)}r_h^{-\frac{3(d-
\theta)}{d}} +\gamma_2\frac{d^4R(d+1-\theta)^2}{4\pi^2 b^2 (d-\theta)^2(4d-3\theta)^2}r_h^{-
\frac{6(d-\theta)}{d}}\right].
\end{eqnarray}
The corrected  Helmholtz free energy is calculated by
\begin{eqnarray}
F&=&\frac{(d+1-\theta) }{4\pi b}\frac{r_h}{\sqrt{1-\left(\frac{r_h}{r}\right)^{d+z-\theta}}}
\left[\frac{8\pi b(d-\theta)(4d-3\theta)}{5d^3 R^5 (d+1-\theta)}r_h^{\frac{3(d-\theta)}{d}}
-\frac{8\alpha}{3R^3} \right.\nonumber\\
&-&\left. \gamma_1\frac{d^2(d+1-\theta)}{2\pi bR(d-\theta)(4d-3\theta)}r_h^{-\frac{3(d-
\theta)}{d}} +\gamma_2\frac{d^4R(d+1-\theta)^2}{4\pi^2 b^2 (d-\theta)^2(4d-3\theta)^2}r_h^{-
\frac{6(d-\theta)}{d}}\right.\nonumber\\
&-&\left. \frac{4\pi b(d-\theta)(4d-3\theta)}{d^{2}R^{2}(d+1-\theta)}r_{h}^{\frac{3(d-\theta)}
{d}}
+\alpha \log \left[\frac{4\pi b(d-\theta)(4d-3\theta)}{d^{2}R^{2}(d+1-\theta)}r_{h}
^{\frac{3(d-\theta)}{d}}\right]\right. \nonumber\\
&+&\left. 2\alpha \log \left[ \frac{(d+1-\theta)r_{h}}{4\pi b R^{3}} \frac{1}{\sqrt{1-
\left(\frac{r_h}{r}\right)^{d+z-\theta}}}\right]-\gamma_1
\frac{d^{2}R^{2}(d+1-\theta)}{4\pi b(d-\theta)(4d-3\theta)}r_{h}^{-\frac{3(d-\theta)}{d}}
\right.\nonumber\\
&-&\left. \gamma_2\frac{d^{4}R^{4}(d+1-\theta)^2}{16\pi^2 b^2(d-\theta)^2(4d-3\theta)^2}
r_{h}^{- \frac{6(d-\theta)}{d}}\right].
\end{eqnarray}

\begin{figure}[h!]
 \begin{center}$
 \begin{array}{cccc}
\includegraphics[width=50 mm]{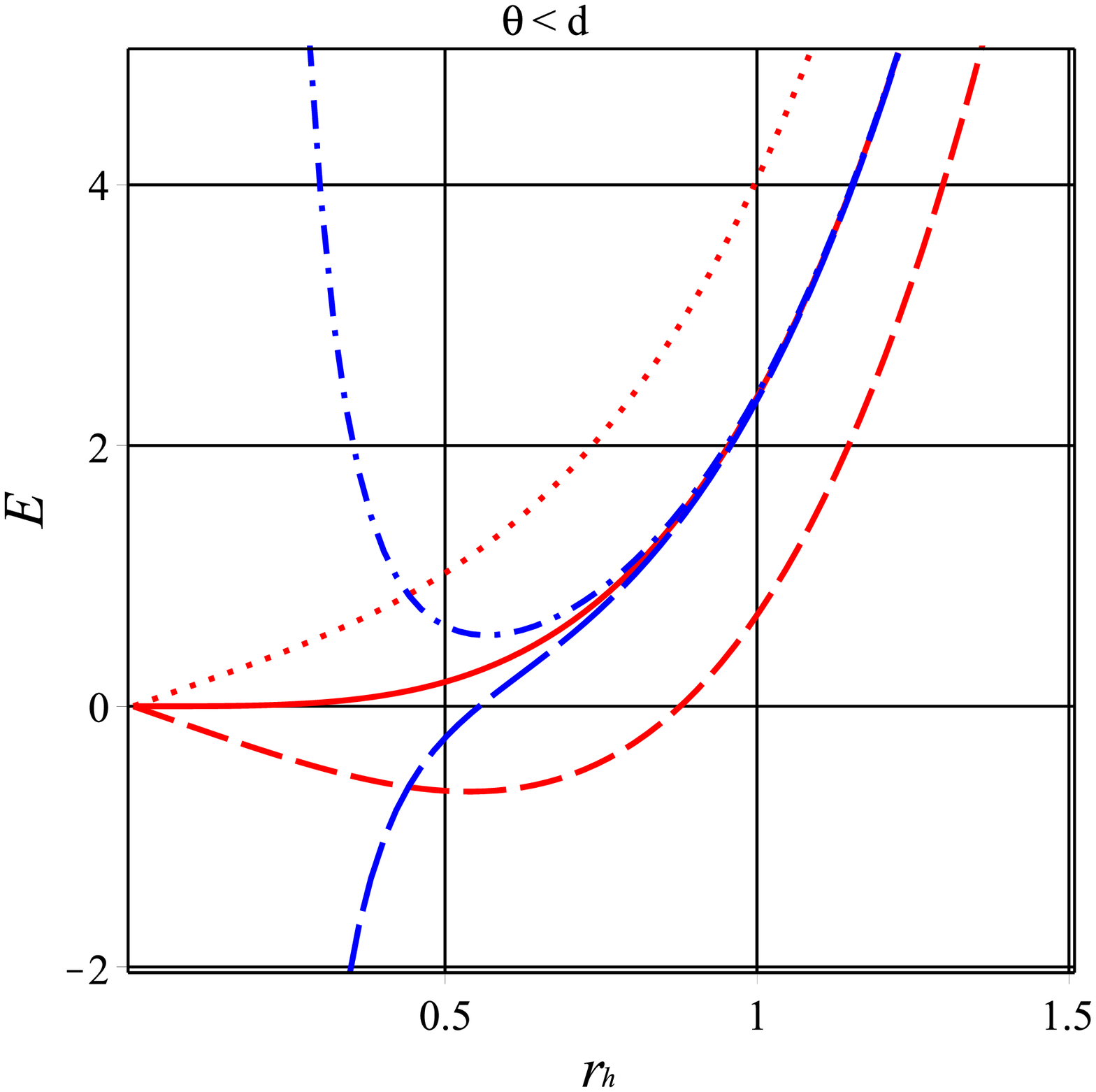}&\includegraphics[width=50 mm]{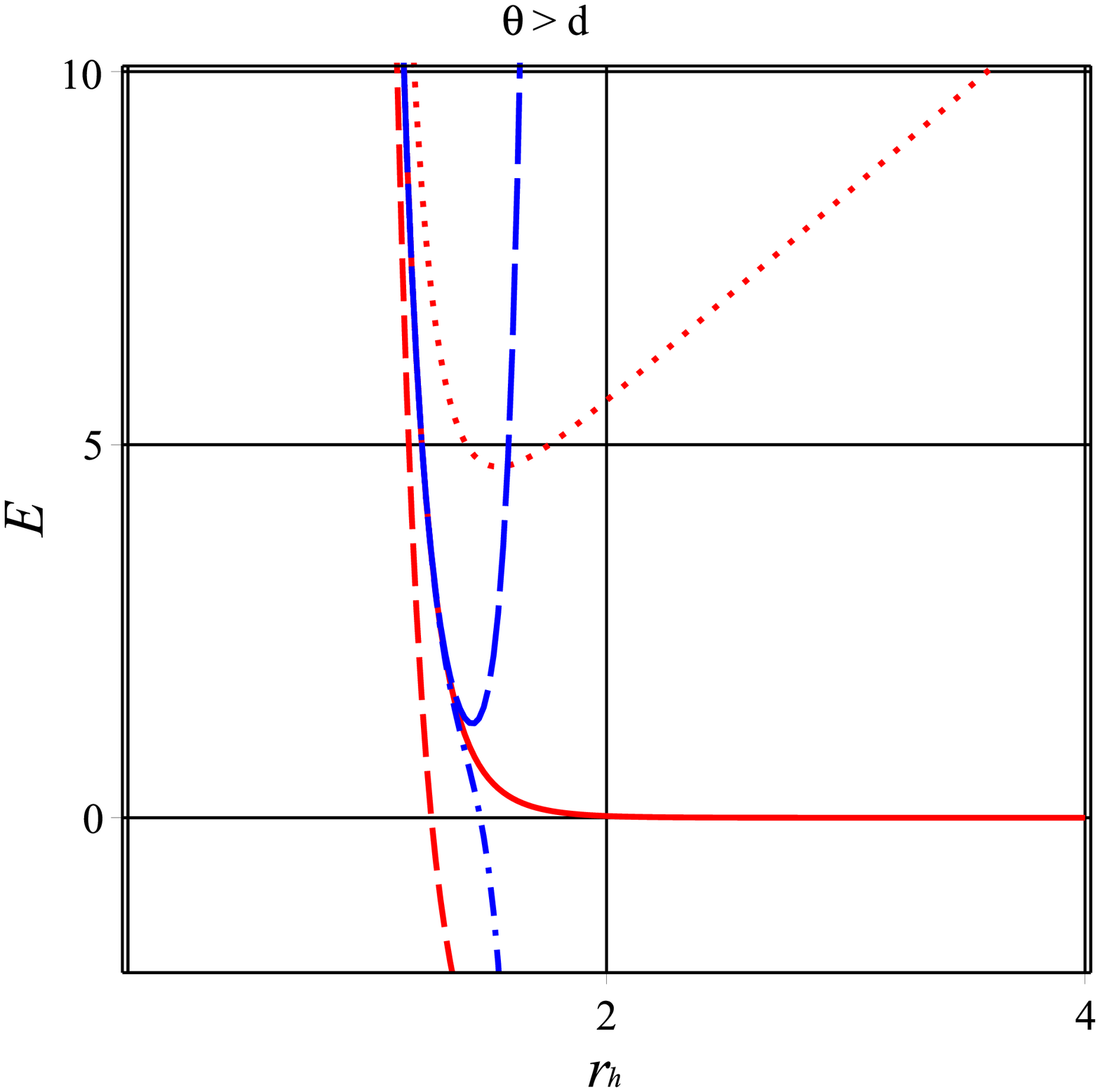}&\includegraphics[width=50 mm]{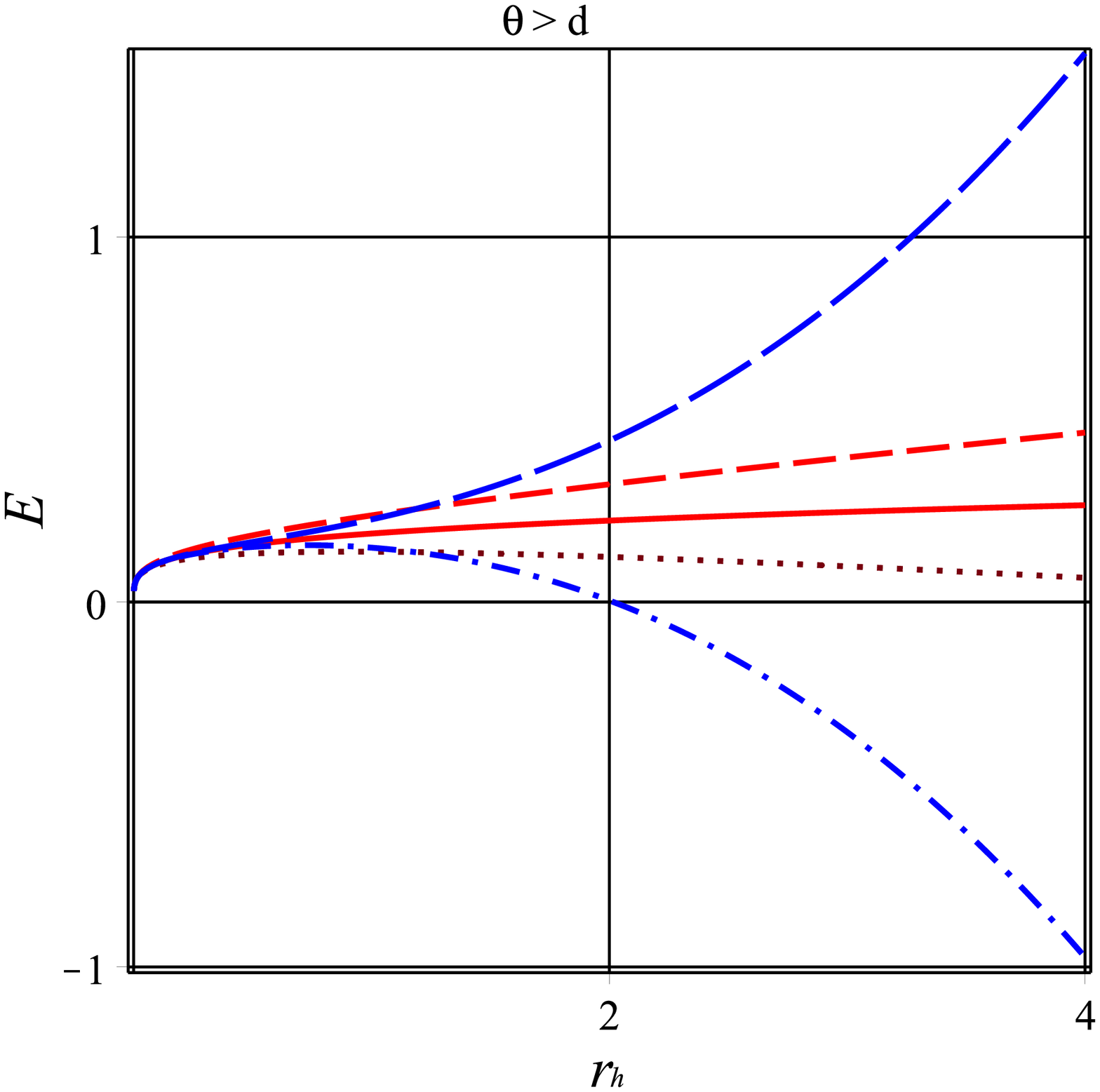}
 \end{array}$
 \end{center}
\caption{Internal energy in terms of $r_{h}$. We set
unit values for all  parameters in this model.
Left: $\theta<d$ for $\theta=1$, $d=9$. Middle: $\theta>d$ for $\theta=10$,
$d=2$. Right: $\theta>d$ for $\theta=10$, $d=8$. $\alpha=0.5$, $\gamma_{1}=0$ (red dash),
$\alpha=-0.5$, $\gamma_{1}=0$ (red dot),$\alpha=0$, $\gamma_{1}=0$ (red solid),
$\alpha=0$, $\gamma_{1}=0.5$ (blue dash dot),
$\alpha=0$, $\gamma_{1}=-0.5$ (blue long dash).}
 \label{fig1}
\end{figure}

In the plots given in the Fig. \ref{fig1},
we plot the  internal energy in terms of horizon radius
for two cases i.e.,  $\theta<d$ and $\theta>d$.
We analyze such a geometry with  $d=2,3,8,9$.
Now, for  $\theta<d$, the behavior of the system does not depend
on the dimension, and the system in all these different dimensions has the same behavior.
However, for $\theta>d$,  there are two phases of the system, i.e.,
$d=2,3$ (middle plot of the Fig. \ref{fig1}), and $d=8,9$
(right plot of the Fig. \ref{fig1}). The
  Helmholtz free energy of this system also has a similar behavior.

\section{Stability}
\begin{figure}[h!]
 \begin{center}$
 \begin{array}{cccc}
\includegraphics[width=75 mm]{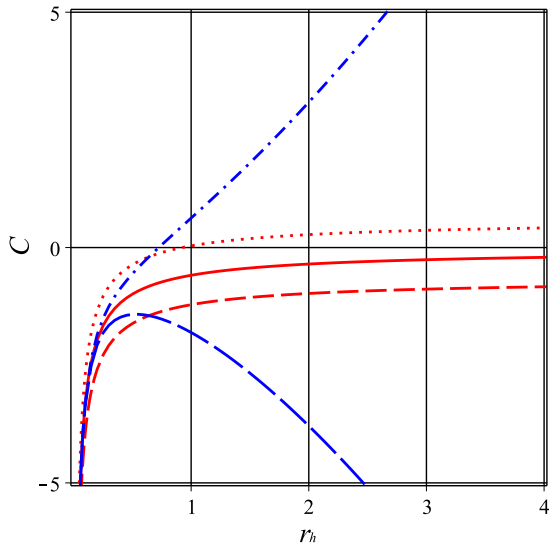}
 \end{array}$
 \end{center}
\caption{Specific heat in terms of $r_{h}$. We set unit value for all
model parameters except $\theta=10$, $d=8$. $\alpha=0.5$, $\gamma_{1}=0$ (red dash),
$\alpha=-0.5$, $\gamma_{1}=0$ (red dot),$\alpha=0$, $\gamma_{1}=0$
(red solid),$\alpha=0$, $\gamma_{1}=0.5$ (blue dash dot),$\alpha=0$,
$\gamma_{1}=-0.5$ (blue long dash).}
 \label{fig2}
\end{figure}

In this section, we will analyze the   stability of this system
using both specific heat and the entire Hessian
matrix of the free energy.   Now, the corrected
specific heat  can be calculated using the corrected entropy of this system,
\begin{equation}\label{16}
C=T\left(\frac{dS}{dT}\right).
\end{equation}
This is given by
\begin{eqnarray}
C&=&\frac{1-\left(\frac{r_h}{r}\right)^{d+z-\theta}}{1+\frac{(d+z-\theta -2)}{2}\left(\frac{r_h}{r}\right)^{d+z-\theta}}\left[\frac{12\pi b(d-\theta)^2(4d-3\theta)}{d^3 R^2(d+1-\theta)} r_h^{\frac{3(d-\theta)}{d}} -\alpha\left(\frac{5d-3\theta}{d}\right)\right.\nonumber\\
&-&\left.\alpha\frac{(d+z-\theta)}{\left[1-\left(\frac{r_h}{r}\right)^{d+z-\theta}\right]}
\left(\frac{r_h}{r}\right)^{d+z-\theta}-3\gamma_1 \frac{dR^2(d+1-\theta)}{4\pi b(4d-3\theta)}
r_h^{-\frac{3(d-\theta)}{d}}\right.\nonumber\\
&-&\left. 3\gamma_2 \frac{d^4R^4(d+1-\theta)^2}{8\pi^2b^2\theta (d-\theta)(4d-3\theta)^2}
r_h^{-\frac{6(d-\theta)}{d}}\right].
\end{eqnarray}
We can analyze the behavior of the
  specific heat for this system using numerical techniques.
We observe  that for  $\theta= {d}/{2}$, there is thermodynamically stable
phase exists for this system
\cite{hyper1}. However, there are phases in this system,
where specific heat is negative. For such phases, it is possible to use the corrected
entropy  to obtain a positive specific heat.
The case of $\theta>d$ is very interesting. In this case,
the specific heat can become  negative when  ($d=8, \theta=10$).
In the Fig. \ref{fig2}, we can see the  behavior of the specific
heat. Solid red line of the Fig. \ref{fig2},
shows that specific heat is completely negative for $d=8$ and $\theta=10$.
It may be noted that   $\alpha=0.5$  produces a negative
specific heat,  while negative value of $\alpha$ produces a stable region. Similar positive
regions can be  obtained by using a positive value of  $\gamma_{1}$
(also for positive $\gamma_{2}$).
We find that infinitesimal positive value of $\alpha$ also produces a negative specific
heat.  Without logarithmic corrections, we can  always obtain positive specific heat
with infinitesimal $\gamma_{1}$. However, we can still obtain instability for the small
black geometries (small $r_{h}$). Hence, as we can see from dash dotted
line of the Fig. \ref{fig2}, corresponding to $\gamma_{1}>0$ and $\alpha=0$,
that there is asymptotic behavior for specific heat which shows unstable/stable
black hole phase transition.  This is different for phase transitions in
hyperscaling-violating geometries    \cite{hyper4}.

\begin{figure}[h!]
 \begin{center}$
 \begin{array}{cccc}
\includegraphics[width=75 mm]{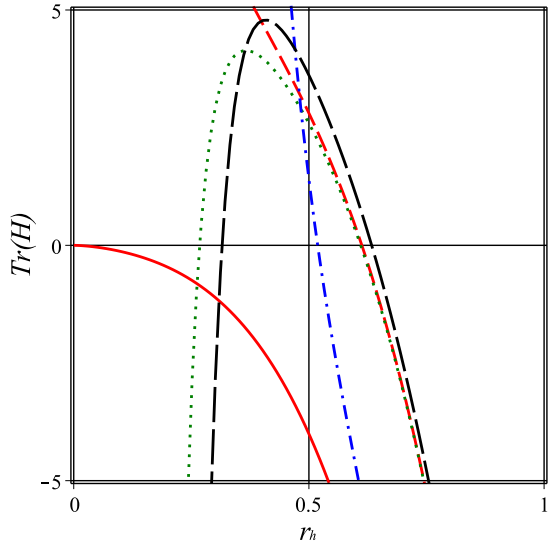}
 \end{array}$
 \end{center}
\caption{Hessian trace in terms of $r_{h}$. We set unit value for
all  parameters in the model  except $\theta=1$, $d=8$. $\alpha=0.5$, $\gamma_{1}=0$ (red dash),
$\alpha=0.5$, $\gamma_{1}=-0.02$ (green dot), $\alpha=0$, $\gamma_{1}=0$ (red solid),
$\alpha=0$, $\gamma_{1}=0.5$ (blue dash dot), $\alpha=0.5$, $\gamma_{1}=-0.1$,
$\gamma_{2}=0.5$ (black long dash).}
 \label{fig3}
\end{figure}

Due to presence of the chemical potential,  we can  analyze this system  using the
matrix of second derivatives of free energy with respect to temperature $T$ and chemical
potential $\mu$, which is given by
\begin{eqnarray}\label{17}
H_{11}= \frac{\partial^{2}F}{\partial T^{2}},&&
H_{12}=\frac{\partial^{2}F}{\partial T\partial\mu},\nonumber\\
H_{21}=\frac{\partial^{2}F}{\partial\mu\partial T},&&
H_{22}= \frac{\partial^{2}F}{\partial \mu^{2}}.
\end{eqnarray}
Now  $H_{11}H_{22}-H_{12}H_{21}=0$, implies   that one of the
eigenvalues is zero, and so we should have to use the other.
This can be expressed
as the trace of the  matrix,
\begin{equation}\label{18}
Tr(H)=H_{11} + H_{22}
\end{equation}
In the Fig. \ref{fig3}, we plot the  variation of Hessian trace
with respect to the  horizon.
Now here we have negative region for $\theta<d$ (specially $\theta=1$, and $d=8$).
By using the higher order corrections, we can now obtain positive regions.
Thus, we can consider  the  positive $\alpha$,
negative $\gamma_{1}$,  and positive $\gamma_{2}$, in this system.

\section{PV-Criticality}
It is possible to study PV-Criticality for a system using the extended phase space
\cite{ext2}. In this formalism, it is possible to define a thermodynamic volume
and pressure
for a geometry \cite{w1, w2}. Then this thermodynamic volume and pressure can be used to study
the critical phenomena for such a geometry. Thus, using the extended phase space
  thermodynamic, the
volume for a    hyperscaling violation background,  can be written as
\begin{equation}
V  = A\, r_h = \frac{\pi  b (4 d-3 \theta ) (d-\theta ) r_h^{4-\frac{3 \theta }{d}}}{d^2 R^2 (d-\theta +1)}
\label{vol}
\end{equation}
The pressure for such a background can be calculated using
\cite{Kubiznak:2012wp}
\begin{equation}
P = \frac{T}{v} ,
\end{equation}
where  $v= 2 \sqrt{ V/\pi}$. So, we can write the pressure for a
hyperscaling violation background as
\begin{equation}
P_0= \frac{ (d-\theta +1)}{8 \pi  b R^3 \sqrt{\frac{b \left(4 d^2-7 d \theta +3 \theta ^2\right) r_h^{4-\frac{3 \theta }{d}}}{d^2 R^2 (d-\theta +1)}}}\frac{r_h}{\sqrt{1-\left(\frac{r_h}{r}\right)^{d+z-\theta}}}. \label{pre}
\end{equation}

\begin{figure}[h!]
 \begin{center}$
 \begin{array}{cccc}
\includegraphics[width=75 mm]{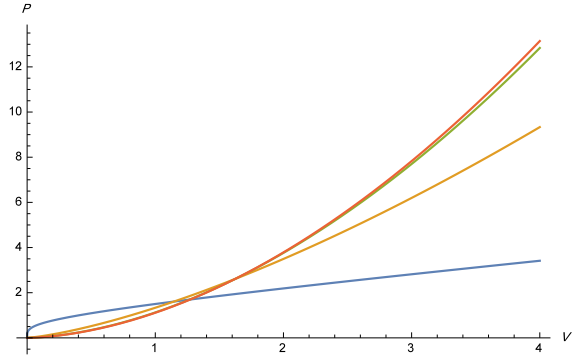}
 \end{array}$
 \end{center}
\caption{P-V diagram for the hyper scaling violating
	geometry for $d=2,3,8$ and $9$. Showing different behavior
	for $d=2$ from the other dimensions. We set $\alpha=0.5$ and $\gamma_{1}=\gamma_{2}=0$.}
 \label{fig4}
\end{figure}

\begin{figure}[h!]
	\begin{center}$
		\begin{array}{cc}
		\includegraphics[scale=0.65 ]{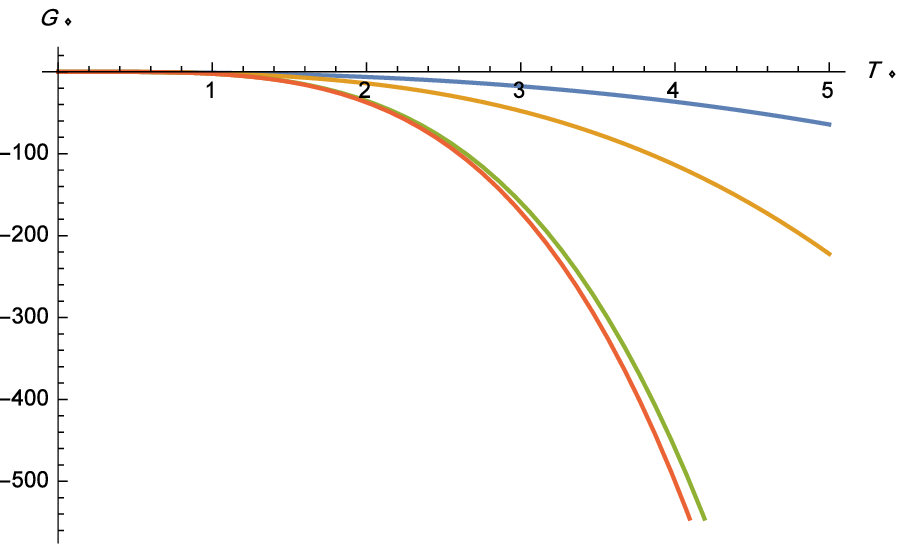}&\includegraphics[scale=0.65]{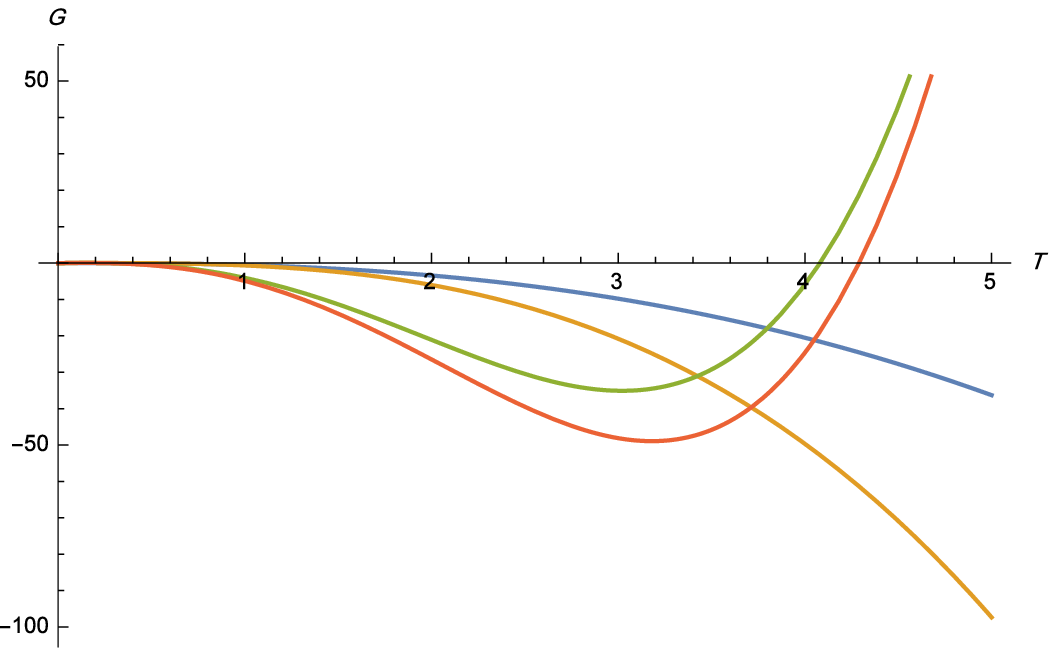}
		\end{array}$
		\caption{ Gibbs free energy as a function of
		temperature, for different space dimensions $ d= 2$ blue,
		$d=3$ yellow, $d=8$ green and $d=9$ red.  On the left,
		we have the Gibbs free energy calculated without thermal
		fluctuations.  On the right,
		thermal fluctuations are considered ($\alpha=0.5, \gamma_{1}=\gamma_{2}=0$).}
	\end{center}
\label{fig5}
	\end{figure}

\begin{figure}[h!]
		\begin{center}
			\includegraphics[scale=0.4]{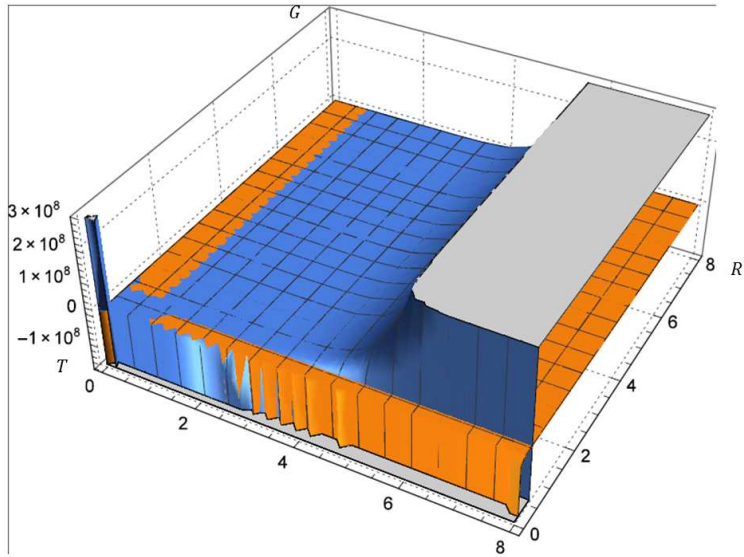}
			\caption{Gibbs free energy $G(T,R)$,
			for $d=3$ in orange and $d=9$ in blue.
			Showing criticality for black hyperscaling
			violating geometries for $d=9$,
			while it there is no criticality
			for the $ d=3$ case ($\alpha=0.5, \gamma_{1}=\gamma_{2}=0$). }
		\end{center}
		\label{fig6}
	\end{figure}

The  Fig. \ref{fig4},  we plot  the PV-diagram for the
hyperscaling violating black geometry (for fixed $R,B$,
and  for different dimensions).
Now, we are able to calculate the  enthalpy  for the system as
\begin{eqnarray}
{\cal H} &=&\frac{(d+1-\theta) }{4\pi b}\frac{r_h}{\sqrt{1-\left(\frac{r_h}{r}\right)^{d+z-\theta}}}
\left[\frac{8\pi b(d-\theta)(4d-3\theta)}{5d^3 R^5 (d+1-\theta)}r_h^{\frac{3(d-\theta)}{d}}
-\frac{8\alpha}{3R^3} \right.\nonumber\\
&-&\left. \gamma_1\frac{d^2(d+1-\theta)}{2\pi bR(d-\theta)(4d-3\theta)}r_h^{-\frac{3(d-\theta)}{d}} +\gamma_2\frac{d^4R(d+1-\theta)^2}{4\pi^2 b^2 (d-\theta)^2(4d-3\theta)^2}r_h^{-\frac{6(d-\theta)}{d}}\right]\nonumber\\
&+&\frac{1}{8R^4 d}\sqrt{\frac{(d+1-\theta)(4d-3\theta)(d-\theta)}{1-\left(\frac{r_h}{r}\right)^{d+z-\theta}}r_h^{6-\frac{3d}{\theta}}}.
\end{eqnarray}
 We may use this  PV relation to study the criticality. This can be done by using
 the Gibbs free energy $ G(T) = E +PV-TS$.
The Gibbs free energy for the hyperscaling violating black geometry  in the extended phase space
can be written as
\begin{equation}
G_0 = \frac{(d-\theta ) (4 d-3 \theta ) r_h^{5-\frac{3 \theta }{d}}}{8 d^2 R^5 \sqrt{\frac{b \left(4 d^2-7 d \theta +3 \theta ^2\right) r_h^{4-\frac{3 \theta }{d}}}{d^2 R^2 (d-\theta +1)}}}-\frac{(4 d-3 \theta ) (d-\theta ) r_h^{\frac{3 (d-\theta )}{d}+1}}{d^2 R^5}+r_h
\end{equation}
It is expected that the thermal fluctuations will correct this expression for the Gibbs
free energy,
and it is possible analyze  the effects of thermal fluctuations on criticality using
this corrected Gibbs free energy. In order to see effects of correction terms,
we set $\alpha=0.5$ and $\gamma_{1}=\gamma_{2}=0$.  The Gibbs free energy corrected by
 thermal fluctuations,    can be written as
\begin{equation}
G = \frac{(d-\theta ) (4 d-3 \theta ) r_h^{5-\frac{3 \theta }{d}}}{8 d^2 R^5 \sqrt{\frac{b \left(4 d^2-7 d \theta +3 \theta ^2\right) r_h^{4-\frac{3 \theta }{d}}}{d^2 R^2 (d-\theta +1)}}}-\frac{(4 d-3 \theta ) (d-\theta ) r_h^{\frac{3 (d-\theta )}{d}+1}}{d^2 R^5}+ d\, r_{h}^{\frac{3 (d-\theta )}{d}+1}.
\end{equation}
It may be noted from  Fig. \ref{fig5}, that there is
 no criticality in any dimension and at any temperature, if thermal fluctuations are neglected.
 In the Fig. \ref{fig6}, we consider logarithmic correction,
and neglect higher order corrections. We analyze the
Gibbs free energy as a function of $T$ and $R$, for $d=3$
and $d=9$.
We can see criticality for black hyperscaling violating geometries
for   ($d=9$), while there is no criticality for the $d=3$. So, the
 Gibbs
free energy also indicates that  the
 criticality of this system  does not change in two and three dimensions due
to thermal fluctuations. However,
for higher dimensions there is a phase transformation at high temperature.
Thus, the thermal fluctuations can change the critical phenomena in such geometries.

\section{Conclusion}
The hyperscaling violating backgrounds are interesting subject
of many recent studies \cite{Hyper010, Hyper011, Hyper012, Hyper013}. In this paper,
we have studied the thermodynamics of
a black geometry with hyperscaling violation.
As this  geometry will be corrected
due to quantum gravitational effects, we expected corrections to the themodynamics
of such black geometries. This is because
 quantum fluctuations in this geometry will produce  thermal fluctuations in the
thermodynamics of this system.  As the form of these corrections is universal,
we have used this universal form to analyze its effects on such a black geometry.
We have discussed   the stability of
this system. This was done by using     specific heat  of this system and the entire Hessian
matrix of the free energy for this system. It was demonstrated that the thermal fluctuations
of this system   can modify the behavior of this system, and this can also
effect the   stability of this system. Finally, we used Gibbs free energy to
analyze criticality for this system.

It may be noted that the thermal fluctuations to the thermodynamics of various geometries
has been recently studied \cite{1503.07418, 1505.02373, 1603.01457, 1605.00924}.
However, most of the work done on the corrected thermodynamics has been done using the
logarithmic
correction term. It is known that the thermodynamics of a black geometry will get corrected
from higher order correction terms \cite{more}. It would be interesting to investigate
such correction terms for these geometries. It would be possible that such correction
terms will modify the stability of such systems. The stability of such system can be studied
by using the specific heat. It is also possible to analyze the stability of such system
using the entire Hessian
matrix of the free energy. It would thus be interesting to perform this study for these
geometries with higher order thermal fluctuations.

\end{document}